\documentclass[twocolumn]{article}
                  \usepackage{hyperref}
                  \usepackage{natbib}
                  \usepackage{amssymb}
                  \usepackage{authblk}

\usepackage{bbm}
\newcommand{\indic}[1]{\mathbbm{1}\left[#1\right]}
\usepackage{caption}
\usepackage{booktabs}
\usepackage[table]{xcolor}
\usepackage{mathtools}
\DeclareMathOperator{\E}{\mathbb{E}}
\DeclarePairedDelimiter{\floor}{\lfloor}{\rfloor}
\newcommand*\dif{\mathop{}\!\mathrm{d}}

\ifdefined \DOUBLEBLIND
\author{}
\date{}
\else
\author{Taylor Pospisil}
\author{Ann B. Lee}
\affil{Department of Statistics \& Data Science \\ Carnegie Mellon University \\ Pittsburgh, PA 15289, USA}
\newcommand{\acks}[1]{\section*{Acknowledgements} #1}
\date{}
\fi

\title{\texttt{(f)RFCDE}: Random Forests for Conditional Density Estimation and Functional Data}
\begin{document}

\maketitle
\begin{abstract}
Random forests is a common non-parametric regression technique which
performs well for mixed-type unordered data and irrelevant features, while
being robust to monotonic variable transformations. Standard random forests, however, do not efficiently handle functional data and runs into a curse-of-dimensionality when presented with high-resolution curves and surfaces. Furthermore, in settings with heteroskedasticity or multimodality, a regression point estimate with standard errors do not fully capture the uncertainty in our
predictions. A more informative quantity is the conditional density 
\(p(y \mid x)\) which describes the full extent of the uncertainty in 
the response $y$ given covariates $x$. In this paper we show how 
random forests can be efficiently leveraged for conditional density estimation, 
functional covariates, and multiple responses without increasing computational complexity.
We provide 
open-source software for all procedures with R and Python versions that call a common 
C++ library.
\end{abstract}

\section{Introduction and Motivation}
\label{sec:orgc25c866}
Conditional density estimation (CDE) is the estimation of the
density \(p(y \mid x)\) where we condition the response \(y\) on
observed covariates \(x\). In a prediction context, CDE provides a more
nuanced accounting of uncertainty than a point estimate or
prediction interval, especially in the presence of heteroskedastic
or multimodal responses. 
Conditional density estimation has proven useful in a range of applications; e.g.,  econometrics \citep{li2007nonparametric}, astronomy \citep{desc1} and likelihood-free inference \citep{izbicki2018abc}.

We extend random forests \citep{breiman2001random} to the more challenging problem of CDE, 
while inheriting the benefits of random forests with respect to 
interpretability, mixed-data types, feature selection,
and data transformations. We take advantage of
the fact that random forests can be viewed as a form of adaptive
nearest-neighbor method with the aggregated tree structures
determining a weighting scheme. This weighting scheme can then be
used to estimate quantities other than the conditional mean; in this
case the {\em entire conditional density}. As shown in Section \ref{sec:org2aa4efb}, random forests can also be adapted to take the {\em functional} nature of curves into account by splitting in the domain of the curves.

Other existing random forest implementations such as
\texttt{quantileregressionForests} \citep{meinshausen2006quantile} and \texttt{trtf}
\citep{hothorn2017transformation} can be used for CDE. However, the first procedure does not alter the splits of the tree, and the second implementation does not scale to large sample sizes.
Neither method handles functional covariates and multiple responses.
In Section \ref{sec:org5590e4a} we show 
that our method achieves lower
CDE loss in competitive computational time for several examples. 


In brief, the main contributions of this paper are

\begin{enumerate}
\item 
Our trees are optimized to minimize a conditional density
estimation loss while remaining computationally efficient; 
hence overcoming the limitations of the usual regression approach
due to heteroskedasticity and multimodality.


\item We provide new procedures and public software (R and Python wrappers to a common C++ library) for joint conditional density estimation and functional
data;  this opens the door for interpretable models and uncertainty quantification for a range of modern applications 
involving different types of complex data from multiple sources.

\end{enumerate}

\section{Method}
\label{sec:orgfea20b0}
To construct our base conditional density estimator we follow the usual random forest
construction with a key modification in the loss function, while retaining algorithms with linear complexity. Here we describe our base algorithm; the extension to functional covariates is described in Section \ref{sec:org2aa4efb}.

At their simplest, random forests are ensembles of regression
trees. Each tree is trained on a bootstrapped sample of the data.
The training process involves recursively partitioning the feature
space through splitting rules taking the form of splitting into the
sets \(\left\{X_{i} \le v\right\}\) and \(\left\{X_{i}> v\right\}\) for a
particular feature \(X_{i}\) and split point \(v\). Once a partition
becomes small enough (controlled by a tuning parameter), it becomes
a leaf node and is no longer partitioned.

For prediction we use the tree structure to calculate weights for
the training data from which we perform a \emph{weighted kernel density
estimate} using ``nearby'' points. This is analogous to the
regression case which would perform a weighted mean.

Borrowing the notation of \cite{breiman2001random} and
\cite{meinshausen2006quantile}, let \(\theta_{t}\) denote the tree structure
for a single tree. Let \(R(x, \theta_{t})\) denote the region of feature
space covered by the leaf node for input \(x\). Then for a new
observation \(x^{*}\) we use \(t\)-th tree to calculate weights for each
training point \(x_{i}\) as

\begin{equation*}
w_{i}(x^{*}, \theta_{t}) = \frac{\indic{X_{i} \in R(x^{*}, \theta_{t})}}{\sum_{i=1}^{n} \indic{X_{i} \in R(x^{*}, \theta_{t})}}.
\end{equation*}

We then aggregate over trees, setting \(w_{i}(x^{*}) = T^{-1}\sum_{t=1}^{T}
   w_{i}(x^{*}, \theta_{t})\). The weights are finally used for the weighted
kernel density estimate

\begin{equation}
\label{eq:orgb5eab9f}
  \widehat{p}(y \mid x^{*}) = \frac{1}{\sum_{i=1}^{n} w_{i}(x^{*})} \sum_{i=1}^{n} w_{i}(x^{*}) K_{h}(Y_{i} - y)
\end{equation}

\noindent where \(K_{h}\) is a kernel function integrating to one. The bandwidth can be selected using plug-in methods or through tuning based upon a validation set. Up to this
point we have the same approach as \cite{meinshausen2006quantile}.

Our departure from the standard random forest algorithm is the
criterion for choosing the splits of the partitioning. In
regression contexts, the splitting variable and split point are
often chosen to minimize the mean-squared error loss. For CDE, we
instead choose splits that minimize a \emph{loss specific to CDE} 
\ifdefined\DOUBLEBLIND
\else 
\citep{izbicki2017converting}
\fi

\begin{equation*}
  L(p, \widehat{p}) = \int \int \left(p(y \mid x) - \widehat{p}(y \mid x)\right)^{2} \dif y \dif P(x).
 \end{equation*}

where $P(x)$ is the marginal distribution of $X$.

This loss is the \(L^{2}\) error for density estimation weighted by the
marginal density of the covariates. To conveniently estimate this
loss we can expand the square and rewrite the loss as

\begin{equation}
\label{eq:org13b95af}
  L(p, \widehat{p}) = \E_{X}\left[\int \widehat{p}^{2}(y \mid X) \dif y\right] - 2 \E_{X, Y}\left[\widehat{p}(Y \mid X)\right] + C_{p}
\end{equation}

\noindent with \(C_{p}\) as a constant which does not depend on \(\widehat{p}\). The
first expectation is with respect to the marginal distribution of
\(X\) and the second with respect to the joint distribution of \(X\)
and \(Y\). We estimate these expectations by their empirical
expectation on observed data.

\begin{figure}[htbp]
\centering
\includegraphics[width=0.5\textwidth,height=0.4\textwidth]{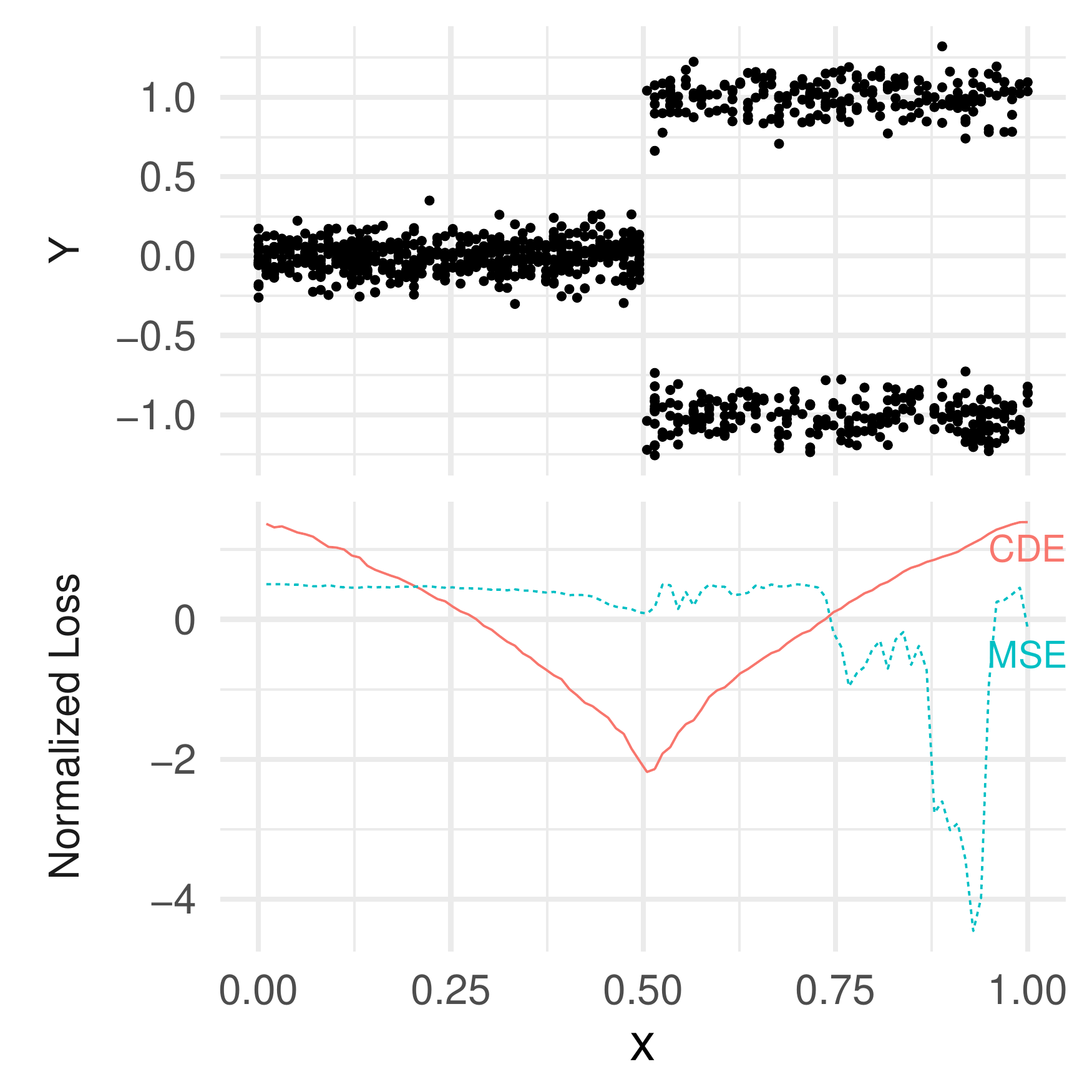}
\caption{\label{fig:org75dcd0e} 
\small (top) Training data with a switch from a unimodal to bimodal response at x = 0.5; (bottom) the normalized losses for each cut-point. MSE overfits to small differences in the bimodal regime while CDE loss is minimized at the true transition point.}
\end{figure}

To provide intuition why switching the loss function is desirable:
consider the example in Figure \ref{fig:org75dcd0e}: we have two
stationary distributions separated by a transition point at
\(x=0.5\): the left is a normal distribution centered at zero and the right is a mixture of two normal distributions centered at 1 and -1. The
clear split for a tree is the transition point: however, because
the generating distribution's conditional mean is constant for all
\(x\) the mean-square-error loss fits to noise and picks a poor split
point far from \(x=0.5\). The CDE loss on the other hand is minimized
at the split point.

While we use kernel density estimates for predictions on new
observations, we do not use kernel density estimates when
evaluating splits: the calculations in Equation \ref{eq:org13b95af} would be expensive for KDE with the term \(\int\widehat{p}(y \mid x)^{2} \dif y\) depending on the \(O(n^{2})\) pairwise distances between all training points.

For fast computations, we instead use \emph{orthogonal series} to compute
density estimates for splitting. Given an orthogonal basis \(\left\{
   \Phi_{j}(y) \right\}\) such as a cosine basis or wavelet basis, we can
express the density as \(\widehat{p}(y \mid x) = \sum_{j}
   \widehat{\beta_{j}}\phi_{j}(y)\) where \(\widehat{\beta_{j}} = \frac{1}{n_{}}
   \sum_{i=1}^{n} \phi_{j}(y_{i})\). This choice is motivated by a convenient formula
for the CDE loss associated with an orthogonal series density
estimate

\begin{equation*}
\label{eq:org3ffb56a}
  \widehat{L}(p, \widehat{p}) - C_{p} = -\sum_{j} \widehat{\beta_{j}}^{2}.
\end{equation*}

The above expression only depends upon the quantities \(\left\{
    \widehat{\beta_{j}} \right\}\) that themselves depend only upon \emph{linear} sums
of \(\phi_{j}(y_{i})\). This makes it computationally efficient to evaluate
the CDE loss for each split.



\section{Results}
\label{sec:org5590e4a}
\subsection{Synthetic Experiment}
\label{sec:orge060632}
\label{org8d1b25b}

To illustrate the potential benefits of optimizing with respect to a CDE loss,
we here compare
against an existing random forest quantile method that minimizes an
MSE loss. More specifically, we adapt the \texttt{quantregForest}
\citep{meinshausen2006quantile} package in R to perform conditional
density estimation according to Equation \ref{eq:orgb5eab9f}. This would then be
 equivalent to our method except that the splits for
\texttt{quantregForest} minimize mean-squared error rather than CDE loss; thereby providing us a means for specifically studying the effect of training random forests for CDE.
We also compare against the \texttt{trtf} package
\citep{hothorn2017transformation} which trains forests for CDE using
flexible parametric families.
\label{orgbf72719}

We generate data from the following model

\vspace{-5mm}
\begin{eqnarray*}
  X_{1:10}, Z_{1:10} \sim \operatorname{Uniform}(0, 1), \quad S \sim \operatorname{Multinomial}(2, \frac{1}{2}) \\
  Y \mid X, Z, S \sim \begin{cases}
    \operatorname{Normal}(\floor*{\sum_{i} X_{i}}, \sigma) & S = 1 \\
    \operatorname{Normal}(-\floor*{\sum_{i} X_{i}}, \sigma) & S = 2 \\
  \end{cases}
\end{eqnarray*}

\noindent where the \(X_{i}\) are the relevant features with the \(Z_{i}\) serving
as irrelevant features. \(S\) is an unobserved feature which
induces multimodality in the conditional densities.

Under the true model \(\E[Y \mid X, Z] = 0\), so there is no
partitioning scheme that can reduce the MSE loss (as the
conditional mean is always zero). As such, the trained
\texttt{quantregForest} trees behave similarly to nearest neighbors as
splits are effectively chosen at random.
We evaluate the three methods on two criterion: training time and
CDE loss on a validation set. All models are tuned with the same
forest parameters (\texttt{mtry} = 4, \texttt{ntrees} = 1000). \texttt{RFCDE} has
\texttt{n\_basis} = 15. \texttt{trtf} is fit using a Bernstein basis of order 5.
\texttt{RFCDE} and \texttt{quantregForest} have bandwidth 0.2.

\begin{table*}[htbp]
\caption{\label{tab:org6c47639}
\small Performance of \texttt{RFCDE} and competing methods on synthetic data; smaller CDE loss implies better estimates.}
\centering
\begin{tabular}{rcccc}
Method & N & CDE Loss (SE) & Train Time (seconds) & Predict Time (seconds)\\
\hline
\texttt{RFCDE} & 1,000 & \textbf{-0.171} (0.004) & 2.31 & 1.29\\
\texttt{quantregForest} & 1,000 & -0.152 (0.003) & \textbf{1.17} & \textbf{0.61}\\
\texttt{trtf} & 1,000 & -0.109 (0.001) & 1013.54 & 123.47\\
\hline
\texttt{RFCDE} & 10,000 & \textbf{-0.194} (0.003) & \textbf{42.99} & 1.95\\
\texttt{quantregForest} & 10,000 & -0.159 (0.003) & 48.60 & \textbf{0.47}\\
\hline
\texttt{RFCDE} & 100,000 & \textbf{-0.227} (0.004) & \textbf{904.76} & 8.05\\
\texttt{quantregForest} & 100,000 & -0.173 (0.003) & 1289.93 & \textbf{0.61}\\
\end{tabular}
\end{table*}

Table \ref{tab:org6c47639} summarizes the results for three simulations
of size 1000, 10,000, and 100,000 training observations. We use 1,000
observations for the test set and calculate the CDE loss. We see
that \texttt{RFCDE} performs substantially better on CDE loss. We also
note that \texttt{RFCDE} has competitive training time especially for
larger data sets. \texttt{trtf} is only run for the smallest data set due
to its high computational cost. 
\subsection{Photo-z Application}
\label{sec:org8ce0da8}
\label{org13487f4}
We next illustrate our methods on applications in astronomy.

In order to utilize the statistical information in images of galaxies, we need to estimate how far away they are from the Milky Way. A metric for this distance is a galaxy's redshift, which is a measure of how much the Universe has expanded since the galaxy emitted its light. For the vast majority of galaxies, estimates of redshift probability density functions are made on the basis of brightness measurements at five wavelengths. This is dubbed {\it photometric redshift} estimation, or {\it photo-z}. Due to degeneracies the pdfs may exhibit, e.g., multi-modality; thus photo-z presents a natural venue for showing the efficacy of\texttt{RFCDE}.


We perform a similar CDE methods comparison as for the synthetic example using realistically simulated photo-z data from LSST DESC\citep{desc1}.
We split 100,000 training observations into subsets of size
1,000, 10,000, and 100,000, and evaluate the CDE loss on 10,000
held-out observations. We compare only against \texttt{quantregForest}, 
dropping \texttt{trtf} again for computational reasons. Both models 
are tuned with the same random forest parameters (\texttt{mtry} = 3, 
\texttt{ntrees} = 1000). \texttt{RFCDE} has \texttt{nbasis} = 31. 
Kernel density bandwidths are selected using plug-in estimates. 
The covariates are the magnitudes for the six LSST filterbands
(ugrizy) together with local differences (u - g, g - r, etc) for
a total of 11 covariates.

\begin{table*}[htbp]
\caption{\label{tab:org40d41fd}
\small Performance of \texttt{RFCDE} and competing methods for photo-z application; smaller CDE loss implies better estimates.}
\centering
\begin{tabular}{rcccc}
Method & N  & CDE Loss (SE) & Train Time (seconds) & Predict Time (seconds)\\
\hline
\texttt{RFCDE} & 1,000 & \textbf{-2.394} (0.033) & 2.24 & 14.53 \\
\texttt{quantregForest} & 1,000  & -2.309 (0.037) & \textbf{1.36} & \textbf{9.85}\\ 
\hline
\texttt{RFCDE} & 10,000 & \textbf{-4.018} (0.054) & \textbf{34.98} & 22.17 \\
\texttt{quantregForest} & 10,000 & -3.809 (0.057) & 40.94 & \textbf{9.54} \\
\hline
\texttt{RFCDE} & 100,000 & \textbf{-5.356} (0.092 ) & \textbf{823.32} & 65.91 \\
\texttt{quantregForest} & 100,000 & -5.064 (0.084) & 2678.39 & \textbf{10.46} \\
\end{tabular}
\end{table*}

Table \ref{tab:org40d41fd} summarizes the results: similarly to the synthetic example
we find that \texttt{RFCDE} achieves substantially better CDE loss with competitive computational 
time.

\subsection{Extension to Joint CDE}
\label{sec:org472ea44}
\label{org7ab3c53}

\texttt{RFCDE} can also target joint
conditional density estimation, which allows us to capture
dependencies in multivariate responses. This feature is
particularly useful for estimating Bayesian posterior distributions in e.g. likelihood free inference \cite{izbicki2018abc}.

The splitting process extends straightforwardly to the multivariate
case through the use of a tensor basis \({p}(y_{1}, y_{2}, \dots, y_{n}
   \mid x) = \sum_{i} \sum_{j} \beta_{i,j}(x) \phi_{j}(y_{i})\) which results in the same
formula for the CDE loss summed over \(\widehat{\beta}_{i,j}^{2}\) instead of
\(\widehat{\beta}_{j}^{2}\).
The density estimation similarly extends through the use of
multivariate kernel density estimation in Equation \ref{eq:orgb5eab9f}.
Bandwidth selection can be treated as in the univariate case
through either plug-in estimators or tuning.


These extensions allow for multivariate CDE estimation (with a current effective limit of three response variables due to the use of the tensor basis). We showcase this by applying \texttt{RFCDE} to a problem related to {\em weak lensing}, the slight perturbation of a galaxy's appearance due to the bending of light by intervening masses. Statistical estimates of how correlated the perturbations are as a function of angular distance allows one to place constraints on two parameters of the Big Bang model: $\Omega_M$, the proportion of the Universe's mass-energy comprised of mass, and $\sigma_8$, which relates to how galaxies have clustered as the Universe has evolved. Specifically, we are interested in inferring the
joint distribution of \(\Omega_M\) and \(\sigma_{8}\) given observed weak lensing data.



\begin{figure}[htbp]
\centering
\includegraphics[width=0.35\textwidth,height=0.35\textwidth]{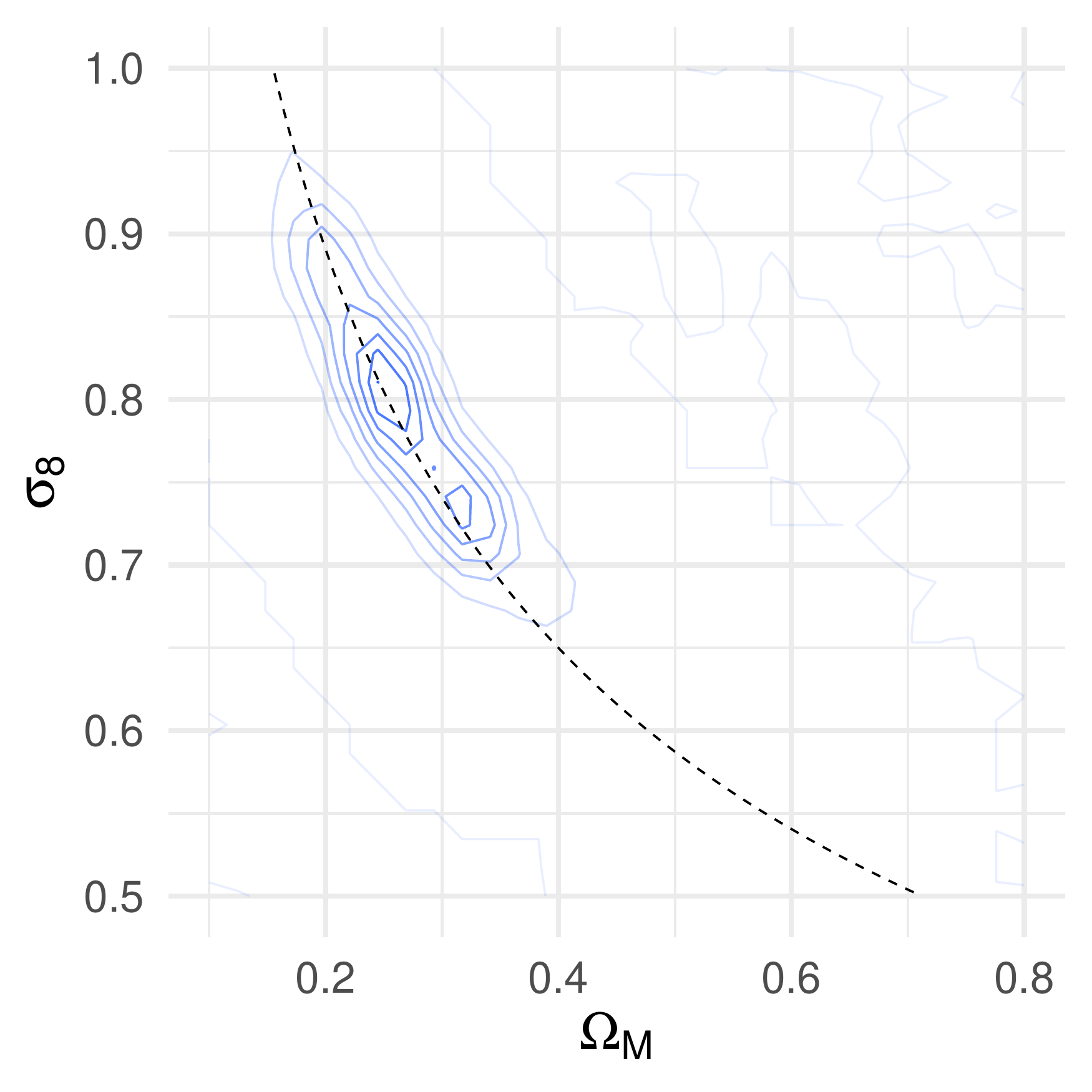}
\caption{\label{fig:org04388cf} 
\small The posterior distribution of the 
parameters $\Omega_{M}$ and $\sigma_{8}$ exhibits a ridge structure which\texttt{RFCDE} captures due to splitting on CDE loss for joint densities.}
\end{figure}

This problem provides an illustrative case
for the need to model joint conditional distributions flexibly. There is a
\emph{degeneracy curve} \(\Omega_{M}^{\alpha}\sigma_{8}\) 
on which the data are indistinguishable.
In Figure \ref{fig:org04388cf} we see that\texttt{RFCDE} can capture this curved ridge structure well which leads to better parameter constraints; in the example we used the \texttt{GalSim} toolkit \citep{rowe2015galsim}
to simulate shear correlation functions under different parameter settings.

\subsection{Extension to Functional Data}
\label{sec:org2aa4efb}
Treating
functional covariates (like correlation functions or images) as unordered multivariate vectors on a grid suffers
from a curse-of-dimensionality: as the resolution of the grid
becomes finer you increase the dimensionality of the data but
add little additional information due to high correlation between
nearby grid points.

To adapt to this setting we follow the approach of
\cite{moller2016random} 
by 
 partitioning the functional
covariates of each tree {\em in their domain},
and passing {\em the mean values of the function for each partition} as inputs to the tree.

The partitioning is governed by the parameter \(\lambda\) of a Poisson process; starting at the first element of the function evaluation vector we take the next \(\operatorname{Poisson}(\lambda)\) evaluations as one interval in the partition. We
then repeat the procedure, starting at the end of that interval and continuing until we have partitioned the entire function domain into disjoint regions $\left\{(l_{i}, h_{i})\right\}$. Once we have partitioned the function domain we take the function mean value $\int_{l_{i}}^{h_{i}} f(x) dx$ within each region as the covariates for the random forest.

We illustrate our functional \texttt{RFCDE} method on spectroscopic data for 3,000 galaxies from the Sloan
Digital Sky Survey. The functional covariates consist of the observed spectrum of each galaxy at 3,421 different wavelengths; the response is the redshift of the galaxy.
These spectra
can be viewed as a high-resolution version of the photometric
data from Section \ref{org13487f4}.

To show the benefits of taking the functional structure of the data into account, 
we compare the results of a base version of\texttt{RFCDE} as described in Section \ref{sec:orgfea20b0} with\texttt{RFCDE} for functional covariates. We set \(\lambda\) = 50. Both trees are otherwise trained
identically: \texttt{ntrees} = 1000, \texttt{n\_basis} = 31, and the bandwidths
are chosen using plug-in estimators. We train on 2000 examples and use the remaining 1000 to evaluate the CDE loss.
\begin{table}[htbp]
\caption{\label{tab:org8abad07} 
\small Performance of RFCDE for functional data. We see that we obtain both better CDE loss and computational time by taking advantage of the structure of the functional data.}
\centering
\begin{tabular}{lrl}
Method & Train Time (sec) & CDE Loss (SE)\\
\hline
Functional & \textbf{12.62} & \textbf{-29.623} (0.844)\\
Vector & 21.31 & -12.578 (0.418)\\
\end{tabular}
\end{table}

Table \ref{tab:org8abad07} shows the CDE loss for both the
vector-based and functional-based RFCDE models on the SDSS data.
{\em We obtain substantial gains with a functional approach}
both in terms of CDE loss as well as computational time. The
computational gains are attributed to requiring fewer searches for
each split point as the default value of \texttt{mtry} = sqrt(d) is
reduced.

\section{Summary}
\label{sec:org04e5b0c}
We adapt random forests to conditional density estimation through
the introduction of an alternative loss function for selecting
splits in the trees of the random forest. This loss function is
sensitive to changes in the distribution of responses, which losses
based upon regression can miss. We exhibit improved performance and
comparable computational speed for a variety of different data examples
including functional data and joint conditional distributions.
We provide a software package \texttt{RFCDE} for
fitting this model consisting of a C++ library with R and Python
wrappers. These packages along with documentation are available at 
\ifdefined\DOUBLEBLIND
Github (link in unblinded version)
\else
\url{https://github.com/tpospisi/rfcde}.
\fi

\acks{We are grateful to Rafael Izbicki and Peter Freeman for helpful discussions and comments on the paper. This work was partially supported by NSF DMS-1520786.}

\bibliography{rfcde-paper}
\end{document}